\documentstyle[twocolumn,epsfig]{article}

\begin{document}

\def\nth{n_{\rm th}}
\def\nobs{n_{\rm obs}}
\def\dmin{d_{\rm min}}
\def\macho{{\sc macho}}
\def\newpage{\vfill\eject}
\def\vs{\vskip 0.2truein}
\def\gnu{\Gamma_\nu}
\def\fnu {{\cal F_\nu}}
\def\mass{m}
\def\lum{{\cal L}}
\def\imf{\xi(\mass)}
\def\ilf{\psi(M)}
\def\msun{M_\odot}
\def\zsun{Z_\odot}
\def\met{[M/H]}
\def\vi{(V-I)}
\def\mtot{M_{\rm tot}}
\def\mhalo{M_{\rm halo}}
\def\pp{\parshape 2 0.0truecm 16.25truecm 2truecm 14.25truecm}
\def\la{\mathrel{\mathpalette\fun <}}
\def\ga{\mathrel{\mathpalette\fun >}}
\def\fun#1#2{\lower3.6pt\vbox{\baselineskip0pt\lineskip.9pt
  \ialign{$\mathsurround=0pt#1\hfil##\hfil$\crcr#2\crcr\sim\crcr}}}
\def\ie{{\it i.e., }}
\def\eg{{\it e.g., }}
\def\etal{{ et al. }}
\def\etalc{{\it et al., }}
\def\kpc{{\rm kpc}}
 \def\Mpc{{\rm Mpc}}
\def\mh{\mass_{\rm H}}
\def\mmax{\mass_{\rm u}}
\def\ml{\mass_{\rm l}}
\def\bc{f_{\rm cmpct}}
\def\br{f_{\rm rd}}
\def\kmsec{{\rm km/sec}}
\def\ibl{{\cal I}(b,l)}
\def\dmax{d_{\rm max}}
\def\dmin{d_{\rm min}}
\def\mbol{M_{\rm bol}}


\bigskip
\centerline{\large \bf Microlensing in the Small Magellanic Cloud:}
\centerline{\large \bf lessons from an {\em N-}body simulation}
\renewcommand{\thefootnote}{\fnsymbol{footnote}}
\centerline{{\bf David S.~Graff}\footnote{\tt graff.25@osu.edu}}
{\it\centerline{The Ohio State University}
\centerline{Departments of Physics and Astronomy}
\centerline{174 W. 18th Ave, Columbus, OH 43201 USA}}
\vskip 0.1in
\centerline{{\bf Lance T. Gardiner}\footnote{\tt ltg@omega.sunmoon.ac.kr}}
{\it\centerline{Sun Moon University}
\centerline{Department of International Education}
\centerline{Tangjeongmyeon, Asankun, Chungnam}
\centerline {Republic of Korea}}
\medskip
\centerline{Keywords: Magellanic Clouds; Gravitational Lensing}

\section*{Abstract}
We analyse an {\em N-}body simulation of the Small Magellanic Cloud (SMC),
that of Gardiner \& Noguchi
(1996) to determine its microlensing statistics.  We find that the
optical depth due to self-lensing in the simulation is low, $0.4\times
10^{-7}$, but still consistent (at the 90 \% level) with that observed
by the EROS and MACHO collaborations.  This low optical depth is due
to the relatively small line of sight thickness of the SMC produced in the
simulation.  The proper
motions and time scales of the simulation are consistent with those
observed assuming a standard mass function for stars in the SMC.  The
time scale distribution from the standard mass function generates a
significant fraction of short time scale events: future self-lensing events
towards the SMC may have the same time scales as events observed
towards the Large Magellanic CLoud (LMC).  Although some debris was stripped
from the SMC during its collision with the LMC about $2\times10^8$ yr ago,
the optical depth of the LMC due to this
debris is low, a few $\times 10^{-9}$, and thus cannot explain the measured
optical depth towards the LMC.

\section{Introduction}
\setcounter{footnote}{0}
\renewcommand{\thefootnote}{\arabic{footnote}}

At the present time, the greatest mystery posed by microlensing is the
cause of the lensing events towards the Large Magellanic Cloud (LMC).
One of the original motivations for microlensing experiments was the
search for halo dark matter constituents such as brown dwarfs.  Based
on the 6-8 events recorded during the first two years of their search,
the {\sc macho} group determined that if these events were caused by
an intervening halo lensing population, this population would account
for a good fraction of the mass of the halo. However, they also
concluded that the mass of the lenses was greater than 0.1 $\msun$,
ruling out a brown dwarf candidate (Alcock \etal 1997a).  White dwarfs
and other stellar remnants have been eliminated by a closed box model
of halo chemical evolution (Gibson \& Mould 1997) and through a more
general model of cosmological chemical evolution (Fields, Freese \&
Graff 1998).  Currently,, there is no really viable candidate for a
halo lensing population for the LMC.

A great limitation of using the LMC to probe the Galactic halo for
microlensing events is that it only samples a single line of
sight.  In response, there have been attempts to look for halo
microlensing along two other lines of sight, towards M31 (Crotts \&
Tomaney 1996; Ansari \etal 1997) and towards the Small Magellanic
Cloud (SMC).  Indeed, two events have so far been detected towards the
SMC.

Both of these events proved to be different from the ensemble of other
events observed towards the LMC.  The first event, MACHO-97-SMC-1 (Alcock \etal
1997c, Palanque-Delabrouille \etal 1998) had a time scale of 120
days, much longer than those of the events seen towards the LMC.  An event of
such long duration should show a measurable parallax deviation from
the standard light curve due to the circular motion of the earth
around the sun, unless the mass of the lens is very large or the lens
is in the SMC.  The lack of a parallax deviation enabled the {\sc
eros} group to show that this event is likely due to a lens within the
SMC, although it may be due to a high mass lens in the halo.
(Palanque-Delabrouille \etal 1998).  The long time scale of this event
also suggested that the lens was drawn from a different distribution
from that of the LMC lenses.

The second observed SMC event, MACHO-98-SMC-1, was also unique. It was
alerted by the {\sc macho} group to be a caustic crossing event caused
by a binary lens (Becker \etal 1998, Alcock \etal 1998).  A caustic
crossing allows an actual resolution of the spatial structure of the
source star.  Detailed observations every few minutes by the {\sc
eros} collaboration (which observed the end of the caustic crossing)
(Afonso \etal 1998) and by the {\sc planet} collaboration (which
observed the peak of the caustic crossing) (Albrow \etal 1999) and
further photometry by {\sc macho/gman} (Alcock \etal 1998) and by {\sc
ogle} (Udalski \etal 1998) allowed all the collaborations to place
limits or solve for the proper motion of the lens.  For this paper, we
will adopt {\sc planet} model 1 with a proper motion of the lens with
respect to the source of 1.26 km s$^{-1}$ kpc$^{-1}$, which gives the best fit to all
the data (A. Gould, private communication).
This proper motion is (as we will confirm) consistent with the lens
being in the SMC; it is an order of magnitude smaller than the proper
motion expected for a lens in the Galactic halo.

To sum up, while the SMC has been tapped as an alternative probe of
the Galactic halo, both of the microlensing events observed so far
towards the SMC are most likely due to lenses inside the SMC.

There have been no precise theoretical predictions of the microlensing
statistics of the SMC because its structure is poorly
understood.  The SMC is known to be far from virial equilibrium
(Staveley-Smith \etal 1997), having suffered from a recent
encounter with the LMC as well as tidal stress from the
Milky Way (Murai \& Fujimoto 1980; Gardiner, Sawa \& Fujimoto 1994).
One can measure the stellar density in the plane of the sky, but there
is no accepted three dimensional model for the distribution of stars
in the SMC (Westerlund 1997).  Mathewson, Ford \& Visvanathan (1986)
examined  Cepheids in the SMC and found a total depth of 32 kpc.
Caldwell \& Coulson (1986) find a total depth for Cepheids in the SMC
of 20 kpc, based on an inclined plane model.  However, for
microlensing, what is important is the line of sight dispersion in distance about
that plane: Caldwell \& Coulson (1986) find a dispersion about the plane
of $0.10 - 0.13$ mag.  Welch \etal (1987) find an rms thickness of
0.12 mag  corresponding to a total width $(4\sigma)$ of 13
kpc.  Welch \etal (1987) suggest that the Mathewson, Ford and
Visvanathan (1986) data suffered from additional unaccounted scatter
in part because of poor phase coverage in their sample: many of the
Cepheids were only observed during luminosity minima.

Since so little is known about the line of sight structure and
transverse velocity distribution of the SMC, previous attempts to determine the
microlensing statistics have relied on simple models of the SMC.
Palanque-Delabrouille \etal (1998) assumed a double exponential SMC
and considered a range of line of sight scale lengths to derive an
optical depth of $1.0 - 1.8 \times 10^{-7}$.  Sahu \& Sahu (1998)
examined a few simple models of the SMC, with a total line of sight
width of $\sim5$kpc to derive an optical depth of $0.5 - 2.5 \times
10^{-7}$
\footnote{There is an error in the determination of the
optical depth in the version of Sahu \& Sahu (1998) on the astro-ph server
which has been corrected in the published version.}
.  Since there is no measurement of the
transverse velocity distribution in the SMC, Palanque-Delabrouille
\etal (1998), Albrow \etal (1999) and Sahu \& Sahu (1998) all assumed
that the transverse velocity distribution should approximately resemble the
line of sight distribution.  However, as the SMC is elongated along
the line of sight, and possibly consists of several distinct components,
this can at best be a rough estimate.

As these models are not accurate in detail, or even in their basic
parameters, they could only provide an
order of magnitude estimate of the optical depth and time scales of
microlensing events.  Since there are only two detected events, no
greater precision was needed to be consistent with the data.  However,
several more events would be detected by existing microlensing
programs if the present event rate of one per year were to continue, and many
more could in principle be detected (Gould 1999).  There is still a
need to make a more accurate prediction of the microlensing optical
depth due to self-lensing in the SMC.

Both the line of sight dispersion and the transverse velocity
dispersion of the SMC will strongly affect the number and character of
the microlensing events due to self-lensing in the SMC.  The optical
depth, the probability that a particular star is being lensed, is
directly proportional to the line of sight depth of the SMC, while the
time scales of the SMC events are in part determined by the transverse
velocity dispersions of the SMC.  By studying an accurate model of the
SMC, we can theoretically predict these quantities.  Conversely,
comparison of the microlensing statistics generated by a model with
the data will provide additional information to constrain the
structure of the SMC.

In several respects, the most realistic models of the SMC are the
result of computer simulations. Such models can provide complete
information on the 3-D structure and kinematics, which is difficult
and in some cases impossible to obtain by other means.  In this paper,
we examine the {\em N-}body simulation by Gardiner \& Noguchi (1996)
(GN). This simulation represents, to date, the only published detailed
self-gravitating simulation of the SMC conducted with {\em N-}body
techniques; previous work (e.g., Murai \& Fujimoto 1980) relied
exclusively on test particle techniques.  This simulation successfully
reproduces a central bar of the SMC due to a {\em spontaneous} bar
instability as well as other features generated as a result of the
tidal interaction of the SMC with the LMC (represented by a fixed
Plummer potential) and the Galaxy (represented by a fixed logarithmic
halo potential).  Such tidal features included a Wing of the SMC, an
inter-Cloud bridge (both formed as a result of an SMC-LMC encounter
$2\times10^8$ yr ago) and the Magellanic Stream (originating at the
penultimate perigalactic passage of the Magellanic Clouds 1.5 Gyr
ago).  Regarding the internal structure of the SMC, the model produces
large-scale features, namely a central bar and tidally induced spiral
arms, which appear to be related to corresponding structures
delineated by observations of Cepheids (Caldwell \& Coulson 1986).
The kinematics of the model correspond to the pattern seen in the
velocity--right ascension plane delineated by HI and young stellar
objects found by Mathewson et al. (1988), showing conclusively that
this pattern is due to a bar structure viewed nearly end-on.  On the
basis of the successful reproduction of many of the observed
characteristics of the SMC, we will conduct a detailed microlensing
analysis of the simulation by GN.

In Section 2, we will examine this simulation to derive a map of the
microlensing optical depth of the SMC from self-lensing, i.e., lensing
of stars in the back of the SMC by stars in the front of the SMC.  We
will show that the optical depth is concentrated toward the centre of
the bar, though some events should still be present towards the Wing.
We will show that the GN simulation has a lower optical depth than the
most likely value found by the {\sc eros} and {\sc macho} experiments
(Palanque-Delabrouille 1998, Alcock \etal 1998).  However, with only
one event in the {\sc eros} analysis and two in the {\sc macho}
analysis, our low optical depth is still consistent with the
experiments at the 90\% confidence level.

In Section 3, we will
calculate the transverse velocities of the lenses with respect to the
moving line of sight and use these velocities to calculate proper
motions and, assuming a typical lens mass function, the time
scales of the lensing events. The time scales and proper
motion distribution in the simulation will be shown to be consistent with those
observed.

The GN simulation also follows debris from the SMC thrown off by the
interaction with the LMC and the Galaxy.  In Section 4, we will show
that this debris does not explain the observed microlensing optical
depth towards the LMC.  We conclude by presenting a discussion and
summary in Sections 5 and 6.

\section{Microlensing Analysis and Optical Depth}
\subsection{Calculation of the Microlensing Statistics}

Our calculations of the microlensing statistics of the SMC are based
on the ``present day'' snapshot of the GN simulation.  The simulation
is set up so that the SMC is initially represented as a disc galaxy
with a halo of equal mass to the disc, and the simulation contains
both ``disc'' and ``halo'' particles.  At the present epoch, the
disc and halo particles are somewhat mixed in spatial distribution,
and the labels refer only to the initial positions of the particles.
Under the assumption that the disc particles are stars and the halo
particles are nonlensing dark matter, we define our principal model to
be one in which only the disc particles participate in lensing.  We
also tested a model in which both halo  and disc particles
participate in lensing, but with one-half of the mass represented by
both types of particles considered to be in non-lensing gas and dark
matter.  The two models gave qualitatively similar results.

We calculate the microlensing statistics of our {\em N-}body simulation by
generating a statistical ensemble of microlensing ``events''.  A
microlensing event is defined to occur when a lensing object lies
within an Einstein radius of the line of sight to a background
``source'' star.  This probability is generally quite low, which is
why microlensing searches require observations of tens of millions of
source stars to find a handful of events.  To boost our statistics, we
artificially increase the Einstein radius of our simulation by a boost
factor $\sqrt{b}$ which increases the probability that a star will be
lensed by a factor $b$.  For each source particle in the simulation,
we count how many lens particles lie within the boosted Einstein
radius.  The microlensing optical depth to a particular source
particle is then just the number of lenses in front of that particle
divided by the boost factor, $\tau=N/b$.  All the microlensing
statistics discussed below were extracted from this ensemble of
events.

\begin{figure}[htb]
\begin{center}
\epsfig{file=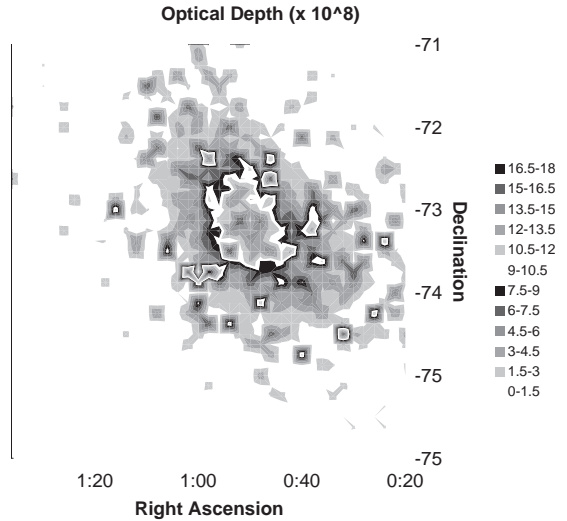,width=9cm}
\end{center}
\vspace{-1.0in}
\caption{\bf The optical depth as a function of position in the SMC.
Isolated spikes are due to discrete
particles.}
\label{tausmcfig}
\end{figure}

\subsection{SMC Optical Depth}
\begin{figure}[htb]
\begin{center}
\epsfig{file=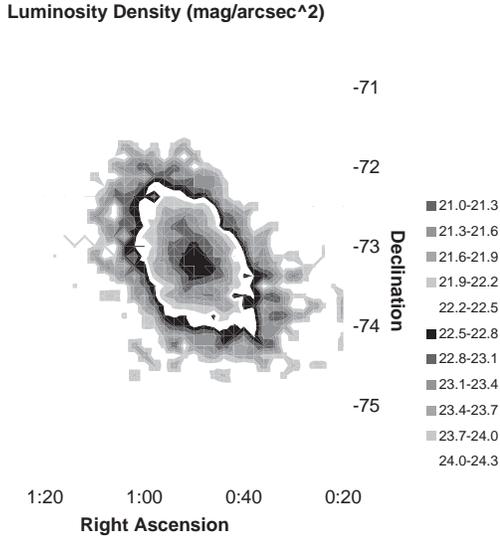,width=9cm}
\end{center}
\vspace{-1.0in}
\caption{\bf The luminosity density of the SMC
from the simulation,  assuming a stellar mass-to-light ratio of 3.
Compare with de Vaucouleurs (1957).}
\label{nmapfig}
\end{figure}

A map of the optical depth of the SMC calculated for the simulation
can be found in Fig. \ref{tausmcfig}.  The optical depth in the centre
of the SMC is relatively high, with a peak of $1.6 \times 10^{-7}$, of
the same order as that measured by the microlensing experiments.

As can be readily seen in Fig. \ref{tausmcfig}, the optical depth of
the SMC is not constant: it is highest in the centre of the bar and
falls off towards the wings.  Comparison with the projected mass
surface density of our simulation shown in Fig. \ref{nmapfig}
shows that in the centre of the bar, the optical depth is roughly
proportional to the projected mass density.  Towards the edges of the
bar, there are some regions with a relatively high optical depth where
there is low mass density; these regions are due to the ``wing''  of
material thrown tens of kpc behind the SMC by the most recent collision with the
LMC.  Isolated peaks in the optical depth in this region are due to
individual particles in the wing lying well behind the SMC.  Particle distribution
plots showing the wing and bar can be found in GN.

However, the microlensing groups do not measure the peak optical depth
of the SMC.  They  report a single number which is an average
optical depth over that portion of the SMC that is observed in their
fields.  In addition, different regions are weighted according to the
local density of source stars in the reference images of the
microlensing surveys, a complex function of the surface brightness,
exposure time, and seeing of the individual experiments.  The
experimental optical depth will thus be
\begin{equation}
\label{tauexpt}
\tau=\frac{1}{N^*}\sum_{\rm resolved \, stars} \hspace{-0.5cm} \tau_i
\end{equation}
where $\tau_i$ is the optical depth of star $i$ and $N^*$ is the total
number of resolved stars in the experiment.  There is an additional
complication due to blending, which is more severe in the inner regions
than the outer ones; however, Monte-Carlo simulations suggest that this
should be at most a 20\% effect (Palanque-Delabrouille 1997).

The optical depth parameter was originally designed to measure the
rate of microlensing due to a foreground Galactic halo population, and was
thought to be a slowly changing function of position.  In that case,
the derived optical depth would be relatively independent of the
averaging, and all experiments should arrive at the same number.
However, as can be seen in Fig.\ref{tausmcfig}, this is not the case:
the SMC self-lensing optical depth rises rapidly to its peak value
roughly as fast as the luminosity density of the SMC rises.  Thus, it
is possible that different experiments averaging over different
fields will report discrepant values of the optical depth.

For example, the {\sc macho} experiment has equal exposure times for
all its fields.  In comparison, the {\sc eros} experiment covers a
wider projected area of the SMC, and has longer exposures and larger numbers
of resolved
stars in the outer fields.  Thus, the {\sc eros} experiment places more
weight on the outer fields with low optical depth than the {\sc macho}
experiment.  We expect the {\sc eros} experiment to report a lower
average optical depth for SMC self-lensing than the {\sc macho}
experiment.

\begin{figure}[htb]
\vspace{-0.5in}
\begin{center}
\epsfig{file=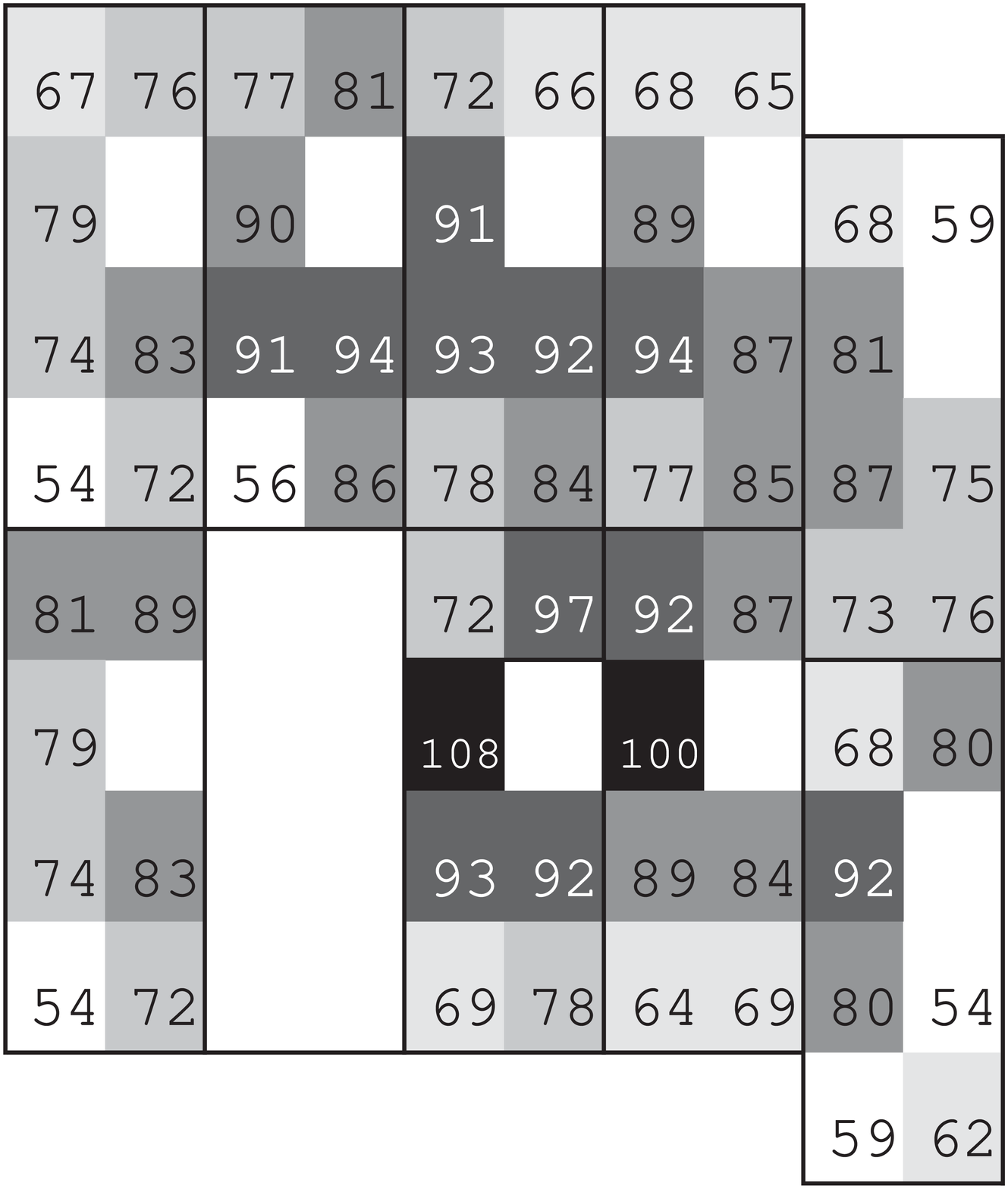,width=9cm}
\end{center}
\vspace{-1.0in}
\caption{\bf The number of source stars in thousands in each CCD in
the {\sc eros} 2 experiment, from Palanque-Delabrouille (1997).  Each
CCD is $0.35^\circ$ on a side.}
\label{sourcefig}
\end{figure}

As an example of an experimental optical depth calculation from the GN simulation, we will determine the mean optical depth that would be
reported by the {\sc eros} experiment with which one of us (DG) has
worked.  The source densities in the {\sc eros} fields can be found in
Palanque-Delabrouille (1997).  We reproduce the number of sources in
each CCD in the {\sc eros} fields in Fig.(\ref{sourcefig}).  We then
approximate equation (\ref{tauexpt}) as an average over the individual
CCDs in the {\sc eros} CCD array:
\begin{equation}
\label{tauccd}
\tau \approx \frac{ \sum_{\rm all\, CCDs} N^*_{\rm CCD} \tau_{\rm CCD}}
                  { \sum_{\rm all\, CCDs} N^*_{\rm CCD} }
\end{equation}
to get $\tau = 0.4\times 10^{-7}$.

If we assume that the density of source stars is equal to the density
of lens stars, we get $\tau = 0.5 \times 10^{-7}$.

\section{Proper Motions and Time Scales}

\begin{figure}[bth]
\begin{center}
\epsfig{file=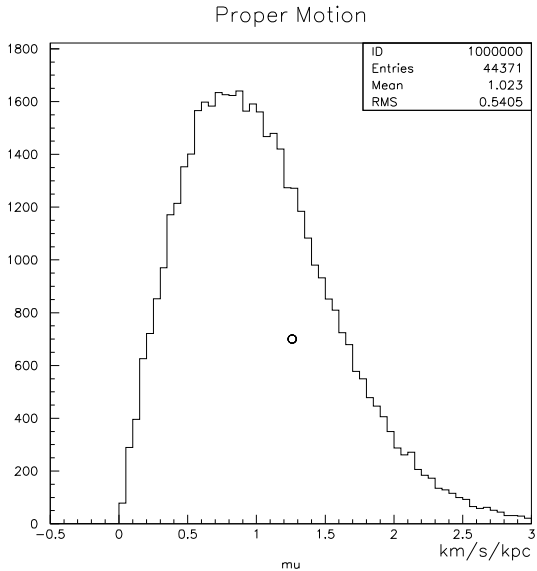,width=8cm}
\end{center}
\caption{\bf Distribution of proper motions of the microlensing events in the
simulation.  The circle marks solution 1 for the proper
motion of MACHO-98-SMC-1 from Albrow \etal (1999).}
\label{mufig}
\end{figure}

The proper motions of our ensemble of SMC events are shown in
Fig.\ref{mufig}.  The typical event in the simulation has a proper
motion of $\sim 1$
km/sec/kpc.  Also shown is the proper motion of the second SMC event.
As can be seen from this figure, the simulation is easily consistent with
the solution for the sole event with an accurately measured proper
motion under the assumption that the lens is in the SMC.

\begin{figure}[bth]
\begin{center}
\epsfig{file=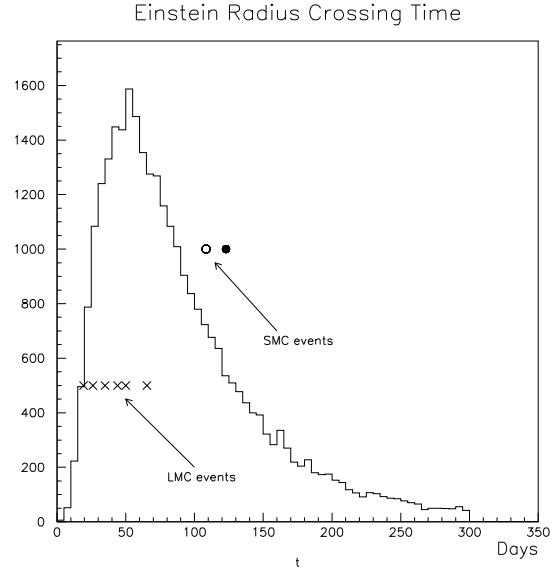,width=8cm}
\end{center}
\caption{\bf Distribution of Einstein radius crossing times of the microlensing
events in the simulation assuming that the lenses have a flat mass
function, $dN/d\log \cal{M}=$const between the cut off masses of
$0.09\msun$ and $1\msun$.  The open circle marks solution 1
for the proper motion of MACHO-98-SMC-1 from Albrow \etal (1999).  The
closed circle is for MACHO-97-SMC-1 from Palanque-Delabrouille \etal
(1998).  The crosses are the LMC events from Alcock \etal (1997a).}
\label{thatfig}
\end{figure}

We calculated the distribution of microlensing time scales (Einstein
radius crossing times) by assuming that the lenses have a flat mass
function, $dN/d\log \cal{M}=$const between the cut off masses of
$0.09\msun$ and $1\msun$.  In this distribution, the average mass of a
lens is 0.35$\msun$.\footnote{The average mass of a lens is different
from the average mass of a star since large mass lenses are more
likely to lens than small mass lenses.}  We find that the mean
Einstein radius crossing time is 100 days while the median
Einstein radius crossing time is 78 days (see Fig.\ref{thatfig}).

Of course, the mass function of the ordinary stellar population of the
SMC is very poorly known, especially at the lower end.  We tried an
alternative mass function, $dN/d\log \cal{M}$ = $m^{-1}$ between the cutoff
masses of 0.01$\msun$ and 1$\msun$.  Note that half of the stellar
mass due to this mass function is in brown dwarfs.  For this model, the
mean Einstein radius crossing time is 50 days while the median time
scale is 30 days.

The long time scales measured in the SMC suggest that the mass
function of the SMC is weighted towards stars of mass 0.3$\msun$, and
is not dominated by brown dwarfs.

\section{LMC Optical Depth}

Zhao (1998) has proposed that debris lying in a tidal tail stripped
from an ancient Magellanic progenitor galaxy by the Milky Way may
explain the observed microlensing rate towards the LMC. Within this
general framework, he suggests that the debris thrown off by the
SMC-LMC tidal interaction could also lead to a high optical depth for
the LMC.  There have been several observational attempts to search for
this debris.  Zaritsky \& Lin (1997) report a possible detection of
such debris in observations of red clump stars, but the results of
further variable star searches by the {\sc macho} group (Alcock \etal
1997b), and examination of the surface brightness contours of the LMC
(Gould 1998) showed that there is no evidence for such a population.
A stellar evolutionary explanation for the observations of Zaritsky \&
Lin (1997) was proposed by Beaulieu \& Sackett (1998).  However,
possible evidence for debris within a few kpc of the LMC along the
line of sight is reported by the {\sc eros} group (Graff \etal in
preparation).

In the GN simulation, there is some debris behind the LMC, but not
enough to explain the observed microlensing events.  The average
optical depth of the LMC due to this background debris is only of order a
few$\times10^{-9}$, two orders of magnitude below that observed.  Even
though the optical depth of the individual stars in the background
debris may be high due to the foreground LMC, the observed optical
depth will be low since the vast majority of observed stars will be in
the foreground LMC, not in the background debris.  Nor is this a matter
of bad luck in the placement of the LMC with respect to the distribution of
LMC debris in the simulation.  Nowhere outside the SMC does the
optical depth due to SMC debris rise above a $\sim$few$\times 10^{-9}$,
orders of magnitude below that observed by the {\sc macho} group.  It
seems unlikely, based on the GN simulation, that the SMC could have
produced enough debris to give rise to a high optical depth for the LMC.

\section{Discussion}

\subsection{Comparison with other results}

The optical depth that we derive for the SMC, $0.4\times 10^{-7}$, is
much lower than the optical depths derived in the simple models of
Palanque-Delabrouille \etal (1998) of $1.0 - 1.8 \times 10^{-7}$ and
Sahu \& Sahu (1998) of $0.5 - 2.5\times 10^{-7}$.  We will examine their
assumptions and show why their optical depths are larger than ours.
The discrepancy arises because the SMC in the GN
simulation has a lower line of sight distance dispersion than was assumed by
Palanque-Delabrouille \etal (1998) and by Sahu \& Sahu (1998).

Palanque-Delabrouille \etal (1998) assume that the density of source
stars follows the density of light whereas in fact it is actually
roughly constant. This is not a bad approximation however, and they
only overestimate the optical depth by a factor of 1.3.  We caution
that if the {\sc eros} fields had been larger, the discrepancy
would have been greater.

Sahu \& Sahu (1998) derive an average mass density by dividing a
fiducial value of the total mass of the SMC of $2\times 10^9 \msun$ by
a fiducial total area of 15 kpc$^2$ to obtain a mass density of
$133\msun/$pc$^2$.  The use of an average mass density is appropriate in
the limit that the source density is constant. This is roughly true for the
{\sc eros} fields which had longer exposures for the outer fields (see
Fig.\ref{sourcefig}), but is likely to be false for the {\sc macho} fields which
had constant exposures.  In fact, given that the {\sc eros} source
stars have roughly unform density, the average mass density should be
computed by dividing the mass within the {\sc eros} fields by the area
of those fields.  The {\sc eros} fields cover an area of about 10
kpc$^2$ and a total mass of about $1\times 10^9 \msun$
(Palanque-Delabrouille \etal 1998) so these estimates are not far off,
yielding a total surface density of $\sim 100 \msun/$pc$^2$, close to
the values derived by Sahu \& Sahu (1998).



\begin{figure}[bth]
\begin{center}
\epsfig{file=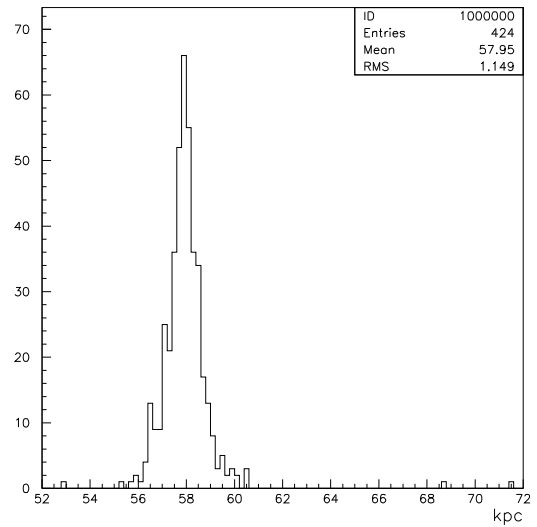,width=8cm}
\end{center}
\caption{\bf Distribution of distances to particles in a small section of the
SMC.
Compare with Fig.9 in GN.  Even though the SMC has a large spatial
extent in the simulation, the extent is due to the large tilt of
the SMC plane with respect to the sky plane; nevertheless the particles are
narrowly distributed about a two dimensional distribution.
}
\label{pencilfig}
\end{figure}

We find that the core of the SMC bar in the GN simulation, where most
of the microlensing takes place in the simulation, is actually rather
narrow, with a line of sight rms dispersion of only 1 kpc (see
Fig.\ref{pencilfig}).  Sahu \& Sahu (1998) and Palanque-Delabrouille
\etal (1998) assume larger dispersions based on the Cepheid
observations of Caldwell \& Coulson (1986) and Mathewson, Ford \&
Visvanathan (1986).  The discrepancy between the simulation and the
Cepheid observations was discussed by GN who suggested that, being a young
population, Cepheids may be influenced by gas-dynamics not present in
the strictly {\em N-}body simulation.  However, since there is no reason to
believe that the lenses in the SMC are due to a particularly young
population, they may in fact be better represented by the {\em N-}body
particles of the simulation than by the (young) Cepheid population.

However, measuring the thickness of a Cepheid population is quite
difficult (Welch \etal 1987).  Caldwell \& Coulson (1986) noted that
their Cepheids were distributed along a plane inclined with respect to
the sky plane; they are not all at the same distance, but neither are
they in a thick distribution liable to generate microlensing.  Graff
\etal (in preparation), in an analysis of the {\sc eros} 2 Cepheid
database, suggest that the extra dispersion in Cepheid magnitudes in
the SMC may be due to a larger intrinsic dispersion for Cepheids in
the SMC compared to the LMC (which had been used as a standard), or
perhaps to larger differential reddening.  The Cepheid samples may be
heavily influenced by Cepheids in the Wing of the SMC.  Although the
Wing does indeed have a large line of sight distance dispersion, it is
not substantial enough in the simulation to contribute strongly to the
microlensing rate.

Our optical depth calculation of $0.4\times 10^{-7}$ is nearly an
order of magnitude lower than that calculated (from only 1 event,
MACHO-97-SMC-1) by the {\sc eros} experiment of $3.3\times 10^{-7}$
(Palanque-Delabrouille \etal 1998) and is also lower than the first
order estimate by the {\sc macho} collaboration of $2-3\times 10^{-7}$
based on both events (Alcock \etal 1998).  Since these optical depth calculations are based
on only a few events, the discrepancy is at most marginally
significant.  The 90\% lower limit on the {\sc eros} optical depth is
only $0.35 \times 10^{-7}$; the corresponding {\sc macho} value is $\sim 0.5
\times 10^{-7}$.

If, however, several more SMC events are detected, thereby confirming
that the optical depth is larger than $\sim 0.4 \times 10^{-7}$, then
we will have learned that the GN model contains deficiencies in its
representation of the SMC.  Either the bar is actually thicker than in
the GN simulation, or the Wing is more substantial.

The length of the initial bar used in the simulation was about 8 kpc;
if the bar had maintained its original configuration to the present day, it could
be more highly inclined to the sky plane while preserving its projected dimensions,
thereby giving a greater thickness along an individual line of sight. A more extended
bar could be realized if the encounter between the Magellanic Clouds a few $10^8$
yr ago was a less
disruptive one. A lower mass for the LMC, a higher mass for the SMC,
and greater separation between the Magellanic Clouds at closest
approach would all serve to reduce the strength of tidal forces
disrupting the edges of the bar. Another possibility to consider is
that intense star formation occurred in the spiral arms generated by
the tidal interaction, contributing to the optical depth towards the
SMC centre.  A hydrodynamical simulation which includes star formation
would be needed to investigate this possibility.

\subsection{Nature of the Lensing Population}

The {\sc eros} group has suggested that measurements of microlensing
in the SMC could be interpreted as showing that there is no foreground halo
population of lenses (Afonso \etal 1998).  They predicted that future SMC
events would continue to have  long time scales as observed in the
first SMC event.  In that case, since SMC events would have different
time scales from those observed in the LMC, they could not be due to
the same foreground halo population.

Our simulation suggests that the situation will not be so clear.  As
seen in Fig.\ref{thatfig}, the predicted range of time scales from a
realistic velocity dispersion and a realistic mass function is broad
enough to encompass short time scale events such as those found in the
LMC (marked as crosses).  Thus, some of the subsequent SMC events should
have the short time scales seen towards the LMC.  We expect that
several SMC events will be needed before one can conclusively say that
events towards the two galaxies come from different distributions.

This result is not specific to our simulation; any reasonably broad
mass function and transverse velocity distribution should give rise to an
equally broad distribution of Eintein radius crossing times.

In addition, if our low estimate of the optical depth of the SMC is
correct, then current experiments will not provide a large enough sample of
future events to analyze in this manner.

The GN simulation does not lend support to Zhao's (1998) suggestion that
tidal debris
from the most recent collision of the SMC and the LMC could be
responsible for the LMC microlensing events. 
Nevertheless, Kunkel \etal
(1997) have suggested that repeated encounters between the SMC and the
LMC could form a polar ring around the LMC; it is feasible that such a structure
might be responsible
for microlensing.  Thus, a simulation with an earlier staring epoch (the
GN simulation only considers the two most recent SMC-LMC encounters) might
slowly build up debris bound to the LMC.   Exploratory
simulations that we have performed with up to four previous encounters do not
support this hypothesis; any nascent ring structure tends to be disrupted
by
tidal interactions with the Milky Way.  However, it is conceivable that the GN
simulation may not accurately track the past history of the SMC-LMC interaction,
and that a different orbital configuration may reproduce a large
enough cloud of debris along the line of sight to the LMC to generate
microlensing.

\subsection{The Future}

If we are correct in claiming that the self-lensing optical depth
towards the SMC is low, then in the near future, present generation
experiments will not detect a large number of self-lensing events.  If
there is a foreground Galactic halo population of lenses, then these
lenses will be detected in significant numbers, and will tend to have
the short time scales seen in the LMC events.

However, if we are incorrect, i.e., the optical depth towards the SMC is
high, then there will be a large background of SMC self-lensing
events.  A good fraction of these events will have the short time
scales seen in the LMC events, and will be indistinguishable from a
putative population of foreground halo lenses.

Ultimately, improved observations (Stubbs 1998) may be necessary to
resolve the question of the nature of lensing events in the Magellanic
Clouds.  Better seeing and larger telescopes will allow the detection of
many more events.  Better photometry will allow the detection of small
changes in the shape of the microlensing light curve which will
provide further information on the location of the lenses.

\section{Summary}

The GN model of the SMC is consistent with the hypothesis that the
two microlensing events observed towards the SMC are due to
self-lensing.  The time scales and proper motions in the model are
consistent with those observed towards the SMC.  The optical depth in
the model, $0.4 \times 10^{-7}$, is much lower than the most likely
value suggested by the experiments, but still consistent at the 90\%
confidence level owing to the paucity of events.  This low optical
depth is due to the low line of sight thickness of the model.  The GN
simulation does not explain the observed microlensing towards the LMC.
Nowhere in the simulation outside the SMC is the debris dense enough
to generate a large optical depth.  Tidal debris from the
most recent SMC-LMC collision is unlikely to account for the LMC
optical depth.

Owing to the low SMC optical depth, suggesting that the number of
future events will be low, and the breadth of the time scale
distribution, which suggests a high probability that at least some of
these future events will have the same time scales as the LMC events,
present generation experiments alone are unlikely to resolve the
question of whether the SMC is lensed by a halo lensing population.

{\it Acknowledgements:} The authors would like to thank Andrew Gould
for useful discussions and a careful reading of the manuscript.
LTG acknowledges the Sun Moon University Research Centre for financial support of
this work.

\section*{Bibliography}
\begin{description}

\item Afonso, C. \etal ({\sc eros}), 1998, A\&A, 337, 17

\item Alcock, C. \etal ({\sc macho}), 1997a, ApJ, 486, 697

\item Alcock, C. \etal ({\sc macho}), 1997b, ApJ, 490, 59

\item Alcock, C. \etal ({\sc macho}), 1997c, ApJ, 491, 11

\item Alcock, C. \etal ({\sc macho/gman}), 1998, submitted to ApJ, astro-ph/9807163

\item Albrow, M. D. \etal ({\sc planet}), 1999, ApJ, in press, astro-ph/9807086

\item Ansari, R. \etal ({\sc eros}), 1996, {A\&A}, {314}, 94.

\item Ansari, R. \etal ({\sc agape}), 1997, A\&A, 324, 843

\item Beaulieu, J.-P., Sackett, P. D., 1998, AJ 116, 209

\item Becker, A. C. \etal ({\sc macho}), 1998, IAU Circ. \#6935

\item Caldwell, J. A. R., Coulson, I. M., 1986, MNRAS, 218, 223

\item Crotts, A. P. S., Tomaney, A. B., 1996, ApJ, 473, 87

\item Fields, B. D., Freese, K., Graff, D. S., 1998, New Astr., 3, 347

\item Gardiner, L. T., Noguchi, M. (GN), 1996, MNRAS, 278, 191

\item Gardiner, L. T., Sawa, T., Fujimoto, M., 1994, MNRAS, 266, 567

\item Gibson, B. K., Mould, J. R., 1997, ApJ 482, 98

\item Gould, A., 1999, submitted to ApJ, astro-ph/9807350

\item Gould, A., 1998, ApJ 499, 728

\item Kunkel, W. E., Demers, S., Irwin, M. J., Albert, L., 1997, ApJ,
488, 129

\item Mathewson, D. S., Ford, V. L., Visvanathan, N., 1986, ApJ, 301, 664

\item Murai, T., Fujimoto, M., 1980, PASJ, 32, 581

\item Palanque-Delabrouille, N., 1997, Ph.D. thesis, Universit\`{e} de
Paris 7 \& University of Chicago.

\item Palanque-Delabrouille, N. \etal ({\sc eros}), 1998,
A\&A, 332, 1, astro-ph/9710194

\item Sahu, K. C., Sahu, M. S., 1999, ApJ {\it in press}, astro-ph/9810053

\item Staveley-Smith, L., Sault, R.J., Hatzimitriou, D., Kesteven, M.J., 
McConnell, D., 1997, MNRAS, 289, 225

\item Udalski, A. \etal ({\sc ogle}), 1998, Acta Astronomica, 48, 383

\item de Vaucouleurs, 1957, AJ, 62, 69

\item Welch, D. L., Maclauren, R. A., Madore, B. F. \& McAlary, C. W., 1987, ApJ, 321,
162

\item Westerlund, B. E., 1997, {\it The Magellanic Clouds}, Cambridge University Press

\item Zaritsky, D., Lin, D. N. C., 1997, AJ, 114, 2545

\item Zhao, H. S., 1998, MNRAS, 294, 139

\end{description}
\end{document}